\documentclass[11pt]{article}
\usepackage{graphicx,float,color}
\usepackage{amsmath}
\usepackage{amssymb}
\usepackage{amsxtra}
\usepackage{hyperref}
\usepackage{comment}

\title{Chemical evolution of antimatter domains 
in early Universe}
\author{A.I.Dembitskaia\\National Research Nuclear University MEPhI, 115409 Moscow, Russia \\demanna2004@yandex.ru\\Stephane Weiss\\National Research Nuclear University MEPhI, 115409 Moscow, Russia\\weiss.stephane62500@gmail.com \\M. Yu. Khlopov\\Virtual Institute of Astroparticle Physics (VIA), \\ 75018, Paris, France, khlopov@apc.in2p3.fr\\M.A.Krasnov \\Institute of Physics, Southern Federal University, 194 Stachki,Rostov-on-Don\\and National Research Nuclear University MEPhI, 115409 Moscow, Russia}
\date{}
\begin{document}
\maketitle

\begin{abstract}
According to modern physics, our Universe is baryon-asymmetric. That phenomenon
can not be described in the frameworks of the Standard Model of particle physics. Globally, the Universe consists of baryon matter. However, some scenarios can lead to the existence of local antimatter
domains.
In the research, the chemical evolution of such an isolated antimatter domain, surrounded by baryonic matter, is studied.
The size of the domain is
estimated according to the conditions of its survival in baryon surrounding, and the process of annihilation at its border is
taken into account.
\end{abstract}

\section{Introduction}\label{s:intro}
Modern concepts of the Universe assume its baryon asymmetry, which means the absence of macroscopic antimatter in an amount comparable to the amount of matter. Nevertheless, due to the the strong nonhomogeniety of baryosynthesis, under certain conditions, local generation of antimatter domains is possible.The standard mechanism of baryosynthesis predicts baryon asymmetry, which might be described by the value, equal to the ratio of density difference between baryons and antibaryons to the density of photons.\cite{Khlopov_2012}
\begin{equation}
    \eta = \frac{n_b - n_{\bar{b}}}{n_\gamma}
\end{equation}
Globally, the Universe is filled with baryonic matter, but there also may be local regions filled with antibaryonic matter (domains dominated by antimatter).\\
The laws of the strong and electromagnetic interaction are the same
for baryons and antibaryons. According to this, we can assume that the evolution of antimatter can be described similarly to the evolution of matter. \\
Despite this, the formation of astronomical
objects that are similar to objects that we know is impossible in the antimatter domain: during the evolution of matter, the products of nucleosynthesis from other stars may enter the region
from the outside. Since the products of nucleosynthesis inside the antistars leave the domain and cannot influence its chemical evolution, the objects inside the domain must have a composition similar to the primary chemical composition formed during the Big Bang. 
It means that the processes happening within the regions of antimatter during its evolution are different from those that happen with matter.
However, in the early Universe
, primary nucleosynthesis processes would occur in the antimatter domain, leading to the formation of antihelium.
The AMS-02 experiment located on the ISS makes it possible to detect antihelium nuclei among the helium nuclei of cosmic rays. If similar results are obtained, it will confirm the possibility of the existence of separate antimatter domains in the universe.
\section{Size of the surviving domain} \label{2}
\begin{figure}[htp]
    \centering
\includegraphics[width=0.8\linewidth]{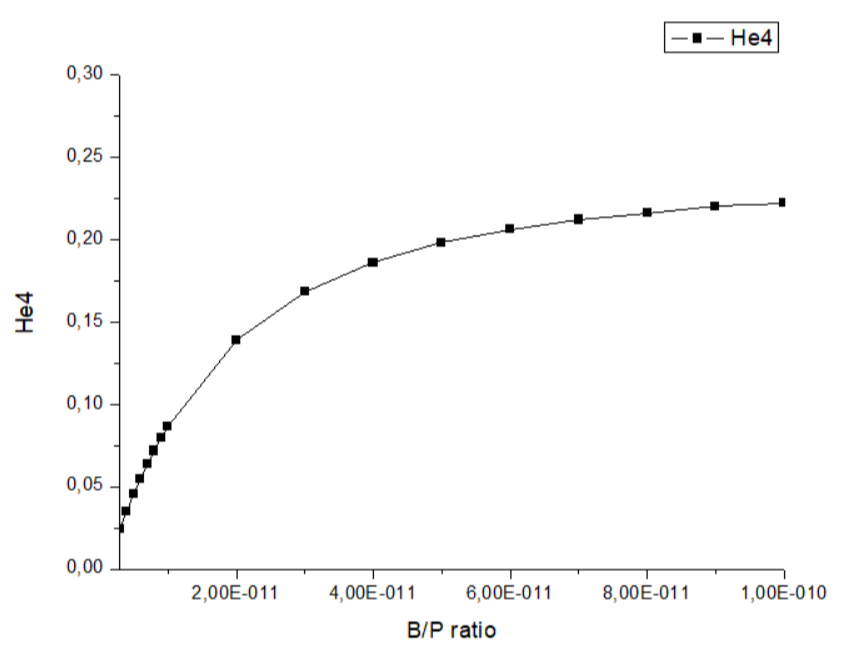}
    \caption{\centering Dependence of the $^4He$ mass fraction on baryon/photon ratio}
    \label{He4}
\end{figure}

\begin{figure}[htp]
    \centering
\includegraphics[width=1\linewidth]{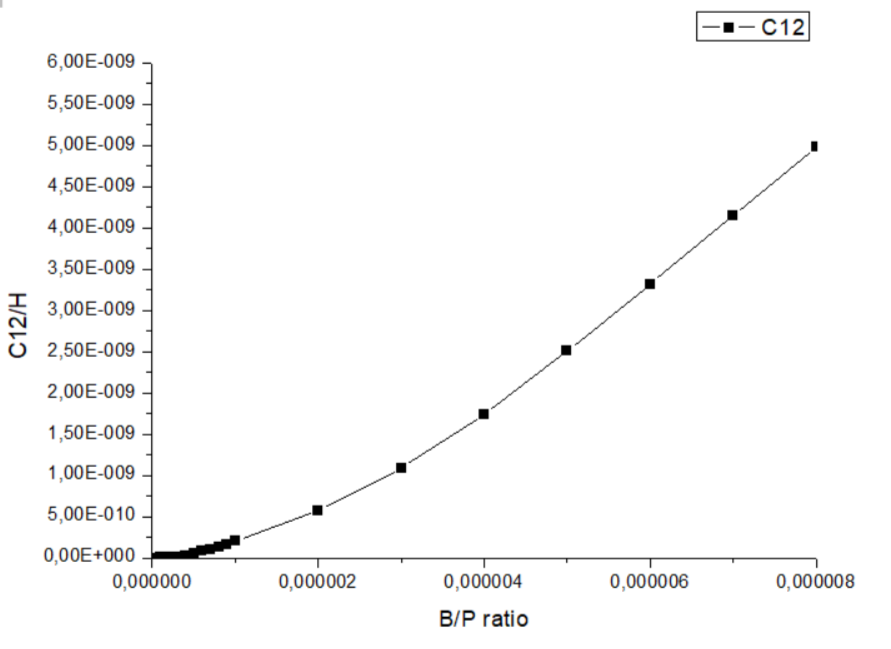}
\caption{\centering Dependence of the $^{12}C$ mass fraction on baryon/photon ratio}
    \label{C12}
    \end{figure}

Since the domain consists of antimatter, during its evolution, annihilation occurs at the boundary of the domain with the horizon. That is why the domain must have a sufficiently large scale to survive to the modern era. Thus, the minimum mass for the domain should be $10^3M_\odot$.\\
It is also necessary that the gamma background should correspond to the observed background\cite{Kneiske_2008}. This constraint defines an upper limit for the mass. Thus, the mass range:\\
\begin{equation}
    10^3M_\odot\leq M\leq 10^5M_\odot.
\end{equation}
Let's assume that the domain we are considering was formed before the era of primary nucleosynthesis, which means that it does not contain heavy elements. The presence of metals in the domain would imply interaction with matter, which would lead to the observed gamma-ray bursts caused by annihilation. Therefore, the domain must be zero-metallic, which leads to certain restrictions on its density. On the one hand, the domain should consist primarily of antihelium, but it cannot contain elements heavier than lithium. \\
The main characteristic of the domain density is the antibaryon-photon ratio. This physical value makes it possible to determine the mass fractions of chemical elements within a domain.\\
To analyze the dependence of the mass fractions of chemical elements on the antibaryon-photon ratio, the AlterBBN program was used.\\
The graphs show the dependence of the mass fraction of the elements formed on the baryon-photon ratio for the following elements: $^{4}\mathrm{He}$ (\ref{He4}), $^{12}\mathrm{C}$ (\ref{C12}). \\
According to the data, the density range of the domain:
\begin{equation} 3\times10^{-12}\leq\eta\leq1\times10^{-6}
\end{equation}

\clearpage
The size of the domain also might be calculated:
\begin{equation}
R=\left(\frac{N}{n}\right)^{\frac{1}{3}},
\end{equation}
where $n=\eta n_\gamma;\\
n_\gamma$-density of thermal photons in the domain.
For the radiation era we have:
\begin{equation}
     \mathrm{R}=\left(\frac{M}{m_p\eta T^3}\right)^{\frac{1}{3}}.
     \label{3}
\end{equation}
Estimated domain size for appropriate temperatures $10^{11}-10^{19}$ sm. Size of the horizon for the same period is $10^{11}-10^{21}$ sm.\\
Therefore, depending on the domain parameters, its size can either exceed the horizon size or be significantly smaller. This will affect the nature of the processes taking place at the domain boundary.\\
Consider the period of time from which the domain becomes smaller than the horizon:
\begin{equation}
    ct\geq R
\end{equation}
Thus, we get time constraints:
\begin{equation}
    t\geq \left(\frac{M}{c^3 m_p\eta 10^{30}}\right)^{\frac{2}{3}}.
\end{equation}
The minimal possible time for such constraints is $t_{min}=1,25\times 10^3$ c which consider to the radiation era. According to the connection between time and temperature at the radiation era, the maximum temperature is $T=2,83\times 10^8\text{K}$. In the future, there is a cooling of the domain associated with the cooling of the Universe.

\section{Antimatter domain in the model of spontaneous baryosynthesis}
One possibility to form antimatter domain is the dynamics of pseudo-Nambu-Goldstone boson (PNGB) arising in a simple model of spontaneous baryogenesis. After phase transition the Lagrangian of the model is as follows:
\begin{multline}\label{AftSymBroken2}
\mathcal{L}=\frac{f^2}{2}\partial_\mu \theta \partial^\mu \theta + i\overline{Q}\gamma^\mu \partial_\mu Q + i\overline{L}\gamma^\mu \partial_\mu L - m_Q \overline{Q}Q - m_L \overline{L}L +\\+
    \frac{gf}{\sqrt{2}}\left(\overline{Q} L e^{i\theta} + \overline{L}Q e^{-i\theta}\right) - V(\theta).
    \end{multline}
In this particular model given by the Lagrangian~\eqref{AftSymBroken2} it was found in \cite{Dolgov_1995, Dolgov_1997} that PNGB rolling from $\pi$ to $0$ produces baryon excess, while rolling from $-\pi$ to $0$ would produce antibaryon excess. Since PNGB's potential posses the following symmetry:
\begin{equation}
    V(\theta)=V(\theta + 2\pi),
\end{equation}
one can conclude that rolling from $\pi$ to $2\pi$ will produce antibaryon excess.
PNGB field would fluctuate during inflation and it could possibly cross $\pi$ which would lead to the formation of domain walls. In such configuration, there would be a region of antibaryon excess inside the wall and a region of baryon excess around the wall.
Size of non-vanishing fluctuations during inflationary stage are determined by Hubble parameter $H_{\text{inf}}$ during that stage. If quantum fluctuation takes place at the moment $t$ during inflation, then by the end of cosmological inflation its size is as follows:
\begin{equation}
    r_{inf}(t)=H_{\text{inf}}^{-1}e^{N_{inf}-H_{\text{inf}}t}.
\end{equation}
After inflation ends these fluctuations would be stretched by consequent expansion. When Hubble parameter $H$ would be of order of $\Lambda^2/f$, which refers to PNGB's mass, classical motion of the PNGB field would start and formation of closed walls would take place.
Mass of the scalar field implies lower limit on size of the antimatter domain, because there are fluctuations which would enter horizon before start of the classical motion of the scalar field. Minimal size is as follows:
\begin{equation}
    r_{hor}=H^{-1}=m^{-1}_{\theta}=\cfrac{f}{\Lambda^2}.
\end{equation}

Then the upper limit on wall's size is determined via its tension. Domain wall with constant surface energy density could start to dominate within Hubble horizon before it could have entered the cosmological horizon if size of this wall is large enough. Following \cite{2016JCAP, Deng_2017}, corresponding timescale at which wall would start to dominate and escape into baby universe is as follows: 
\begin{equation}
    t_{\sigma}=\cfrac{M^2_{Pl}}{2\pi \sigma},
\end{equation}
where $\sigma$ is surface energy density of the wall. In case of considered PNGB it is calculated via model's parameters as follows:
\begin{equation}
    \sigma=4\Lambda^2 f \rightarrow t_{\sigma}=\cfrac{M^2_{Pl}}{8\pi \Lambda^2 f}.
\end{equation}
Combining expressions above, one can estimate threshold for size of antimatter domain $r_{domain}$ as follows:
\begin{equation}
    \cfrac{f}{\Lambda^2} < r_{domain} < \cfrac{M^2_{Pl}}{8\pi \Lambda^2 f}.
\end{equation}

Let us now define a threshold for the mass of the antimatter domain $M$ using estimations of its size above. Let $m_0$ be the mass of a baryon, $n_i$ be the number density of baryons at the moment of wall's crossing the Hubble horizon, then mass of the domain could be estimated as follows:
\begin{equation}
    \cfrac{4}{3}\pi m_0 n_i \left( \cfrac{f}{\Lambda^2}\right)^3 < M < \cfrac{4}{3}\pi m_0 n_i \left( \cfrac{M^2_{Pl}}{8\pi \Lambda^2 f}\right)^3
\end{equation}

\section{Diffusion towards the border of domain}
The phenomenon of diffusion of baryons and antibaryons towards the boundary is governed by elastic scattering, rather than annihilation. The main distinction is important, first of all the elastic scattering processes, which involves interactions where particles maintain their identity but exchange momentum and energy and they are responsible for randomizing trajectories of particle and make easier macroscopic transport of baryons and antibaryons through the primordial plasma. However, annihilation process is a local process that occurs predominantly at domain boundaries where matter and antimatter and in contact and convert into other particles.\\
 The macroscopic transport is achieved by specific elastic scattering mechanisms 

The effective transport of charged anti-baryons in the early Universe is constrained by their coupling to the ambient electron-photon fluid. The Diffusion phenomenon 

\begin{itemize}
    
    \item Proton Electron Elastic Scattering\\
    The direct electromagnetic coupling between charged anti-baryons and the ambient plasma electrons $e^{\pm}$. The process is Coulomb Scattering that is mediated by a virtual photon. To preserve the local charge neutrality, any motion of anti-baryons must be following a motion of electrons.

    \item Electron-Photon Coupling\\
    The mobility of electrons, that limits the transport of anti-baryons is constrained by by their Elastic Scattering against the dense background of thermal radiation (CMB).The interaction $\gamma + e^{\pm} \rightarrow \gamma + e^{\pm}$ creates radiative friction. This has a consequence that the domination constraint on baryon diffusion comes from the Electron-Photon coupling.

    \item Direct Proton-Photon Scattering\\
    Direct Proton-Photon Scattering is negligible due to the extremely low cross-section involved for this interaction.

\end{itemize}

\subsection{Time Depend Evolution of the Diffusion Coefficient}
The Diffusion Coefficient $D$  quantifies the relation between flux of particles and gradient of density. $D$ is proportional to the mean free path $\lambda$ and particle velocity $v_{\text{th}}$. The dependence of time of Diffusion Coefficient depend on the cosmological expansion $a(t)$, which determines the dilution of of the scatterer density $n \propto  a(t)^{-3}$.

In the Radiation-Dominated era, the scale factor evolves as as $a(t) \propto t^{\frac{1}{2}}$. The medium is a relativistic plasma, the velocity is approximately equal to the speed of light $v_{\text{th}} \approx c$. The dilution of the scattered density $n_{\text{e}} \propto t^{-\frac{3}{2}}$ leads to the mean free path $\lambda \propto t^{\frac{3}{2}}$.
\begin{equation}
    D_{\text{rad}} \propto \lambda \cdot c \propto t^{\frac{3}{2}}
\end{equation}
The consequence is the Diffusion becomes more efficient as the Universe expands and the plasma becomes more transparent. $D_{\text{rad}}$ describes the transport lead by the coupling of charged antibaryons with the plasma.

The dynamics change after Radiation-Matter equality when $t>t_{\text{eq}}$. Compare to the Radiation-Dominated era, the scale factor becomes $a(t) \propto t^{\frac{2}{3}}$. However, the recombination $t_{\text{rec}}$ who has a redshift  $z \approx 1100$ hasn't yet occurred, so the charge antibaryons are still interfered with the thermal radiation pressure. The diffusion mechanism remains Radiative Diffusion. The dilution of the scattered density $n_{\text{e}}$ follows a law $n_{\text{e}} \propto a(t)^{-3}$ and because the relation between mean free path and the dilution if scattered density is known 
\begin{equation}
    \lambda \propto \frac{1}{n_{\text{e}}} \propto a^{3} \propto \left (t^{\frac{2}{3}} \right )^{3} \propto t^{2}
\end{equation}
The Diffusion coefficient is also determined by velocity of electrons moving towards the border. The motion is determined by photon pressure and thermal radiation effect. Because of the thermal equilibrium of electrons with plasma we can assume that the velocity could not be equal to the speed of light anymore. So that value also should be estimated as thermal velocity.
\begin{equation}
    v_{th}=\sqrt{\frac{3kT}{m}} \propto t^{-\frac{1}{3}}
\end{equation}
Using the relation between the time and temperature we can estimate the dependence of the Diffusion Coefficient on time as
\begin{equation}
    D_{\text{plasma}}\propto\lambda \cdot v_{th}\propto t^{\frac{5}{3}}
\end{equation}

The mechanism change fundamentally at $t_{\text{rec}}$. Neutralization of antibaryons into antiatoms occurs. Also the electromagnetic coupling with CMD photons disappears, and radiative friction ceases. The transport is governed now by Atomic Diffusion via kinetic collision between neutral atoms.
Th Diffusion Coefficient of kinetic atomic diffusion is given by the fundamental relation of kinetic theory of gases.
\begin{equation}
    D_{\text{atoms}} \propto \lambda_{\text{coll}} \cdot v_{\text{th}}  \propto t^{\frac{4}{3}}
\end{equation}

The Diffusion process is not static but modulated by feedback mechanism from Annihilation. The products of Annihilation $\pi^{0}, \pi^{\pm}$ decay into high-energy photons $\pi^{0} \rightarrow2 \gamma$. The total annihilation energy splits into three components
\begin{itemize}
    \item neutrinos for $50\%$
    \item $\gamma$-photons for $34\%$
    \item $e^{\pm}$ pairs for $16\%$
\end{itemize}
These highly energetic $e^{\pm}$ pairs deposit energy into the surrounding medium, generating a local, non-uniform Annihilation Pressure $P_{\text{ann}}$
\begin{equation}
    P_{\text{ann}} \propto \tau_{\text{rad}} \epsilon \langle \sigma v \rangle n_{\bar{b}} n_{\text{b}}   
\end{equation}
The gradient of the Annihilation Pressure $P_{\text{ann}}$ acts as a drift motion that suppresses the anti-baryon flux towards the interface. 
The total anti-baryon flux $J_{\text{antibaryons}}$ is modeled through the Generalized Fick's Law, which includes both the standard Diffusion term  and the Pressure gradient term
\begin{equation}
    J_{\text{antibaryons}} = -D_{\text{rad}}\nabla n_{\bar{b}} - \frac{D_{\text{rad}}}{k_{\text{B}}Tn_{\text{eff}}}n_{\bar{b}} \nabla P_{\text{ann}}
\end{equation}
To relate this generalized flux to a simplified form 
\begin{equation}
    J_{\text{antibaryons}} = -D_{\text{eff}} \nabla n_{\bar{b}}
\end{equation}
the Thin-Boundary Approximation is applied.
By substituting the gradients with magnitude ration $\nabla n_{\bar{b}} \approx \frac{n_{\bar{b}}}{\delta}$ and $\nabla P_{\text{ann}} \approx \frac{P_{\text{ann}}}{\delta}$, the effective Diffusion Coefficient for Radiation-Dominated era is obtained
\begin{equation}
    D_{\text{eff}} = D_{\text{rad}} \left [1 + \frac{P_{\text{ann}}}{k_{\text{B}}Tn_{\text{eff}}} \right ]
\end{equation}
In the center of the domain, the Annhilation pressure $P_{\text{ann}}$ is vanished, resulting that $D_{\text{center}} \approx D_{\text{rad}}$. This expression is true for Matter Dominated neutral phase. 

\subsection{Processes at the border} \label{3}
Starting from the moment when the size of the horizon exceeds the size of the domain, the annihilation of matter with antimatter will occur at the boundary of the domain, as a result of which high-energy photons will be formed. They will penetrate the domain. 
Since the domain consists entirely of antimatter, annihilation of baryons with antibaryons will occur at the boundary of the region with the horizon, as a result of which various particles will be formed.\\
Consider proton-antiproton annihilation. The cross section of the interaction of this reaction can be described using the experimental data obtained\cite{Golubkov_1999}.
\begin{equation}
\sigma_0\approx1,6\times10^{-25}\text{cm}^2
\end{equation}
This reaction passes through various channels. The probability of each of them can be described using the branching coefficient.
The most probable channels are those involving the formation of neutral and charged pions\cite{Adiels_1987}.\\
The domain that we are considering consists of antimatter and contains positrons inside. As a result of the processes, it is possible for electrons to enter the domain or to be produced it. Then, positron-electron annihilation inside the domain is possible.\\
There are 2 sources of electrons inside the domain:\\
$\cdot$ pair production as a result of interaction between the annihilation and thermal photon;\\
$\cdot$ formation of an electron after the decay of a negatively charged muon, which is the product of a charged pion decay.\\
The result of this reaction will be the formation of annihilation photons inside the domain. \\
After the decay of a neutral pion with an energy of 135MeV, 2 photons are formed, the average energy of each of which is 67,5MeV. \\
Penetrating inside the domain, a high-energy photon can interact with a thermal photon to form a positron-electron pair.\\
The distribution of thermal photons obey the Planck distribution. That guarantees the presence of a non-zero concentration of high-energy photons.If the photon energy is greater than 3,9kev, then the pair-production is possible. Required temperature for that is $T=10^7$K.\\
The interaction of two photons is described by the Breit-Wheeler formula for high energies:
\begin{equation}
\sigma_{pp}\approx\frac{\pi \alpha^2}{s}\left[2\ln\left(\frac{s}{m_e^2}\right)-1\right]
\end{equation}
The energy of a thermal photon is significantly less than the energy of an annihilation photon. In this case, an energy asymmetry is observed and the formula describing the reaction cross section will look like this\cite{Berestetskii_1982}:
\begin{equation}
\sigma_{pp}\approx\frac{2\pi \alpha^2}{s}\left[2\ln{\frac{s}{m_e^2}}-\frac{3}{2}\right]\approx2,1\times10^{-28}\text{cm}^2,
\end{equation}
where $E_1>>E_2$.\\
As a result of the reaction, an electron-positron pair is formed. Moreover, one of the particles will receive almost all the energy of the annihilation photon, while the second will acquire energy comparable to the rest energy of an electron.As the resukt, the annihilation of positrons located in the domain with the resulting high-energy electron is possible.\\
In addition to the formation of electron-positron pairs within the domain, Compton scattering of a photon by a positron is also possible.This process will be dominant at the temperature $T<10^7$K.\\
The relative energy loss after one Compton scattering is described by the following formula:
\begin{equation}
    \frac{\Delta E}{E}=\frac{E'-E}{E}=-1+\frac{1}{1+\epsilon(1-cos\theta)},
    \end{equation}
where $\epsilon=\frac{E}{m_ec^2}$\\
After a single scattering the photon does not lose all its energy. The evolution of the photon distribution during multiple scattering is described by the Kompaneyets equation\cite{Rybicki_1979}:
\begin{equation}
    \frac{\partial n}{\partial y}=\frac{1}{x^2}\frac{\partial}{\partial x}\left[x^4\left(\frac{\partial n}{\partial x}+n+n^2\right)\right],
\end{equation}
where\\
$n$-the number of photons in a state with dimensionless energy $x$;\\
$y=\int \frac{kT_e}{m_ec^2}\sigma_\tau n_pc\mathrm{dt}$-Compton parameter.\\
It follows that the change in photon energy is exponential:
\begin{equation}
    E=E_0e^{-4y}
\end{equation}
The Compton parameter is related to the amount of scattering processes:
\begin{equation}
    y=\frac{kT_e}{m_ec^2}N
\end{equation}
For our temperature range the number of scattering processes required to reduce the photon energy from $E_0=67.5$MeV to $E_N=1$MeV is $N>2,5\times10^3$.
According to the Klein-Nishina formula:
\begin{equation}
    \sigma_{KN}=\sigma_\tau f(x),
\end{equation}
where \\
$\sigma_\tau=\frac{8\pi r_e^2}{3}=6,7\times10^{-25}\text{cm}^2$-Thomson cross section,\\
\hspace{2cm}
$f(x)$-correction factor that takes into account relativistic effects,\\
$x=\frac{E}{m_ec^2}$-the dimensionless energy of a photon.\\
As the photon energy decreases, the cross-section value will approach the Thompson cross-section. \\
Taking into account the numerical solutions of the Kompaynets equation\cite{Rybicki_1979}:
\begin{equation}
    <\sigma>=\sigma_\tau\left[1-\frac{1}{2}\left(1-\frac{m_ec^2}{E_0}\right)\right]\approx0,5\sigma_\tau\approx3,4\times10^{-25}\text{cm}^2
\end{equation}
\begin{equation}
    \sigma_{eff}=N<\sigma>\geq8.4\times10^{-22}\text{ cm}^2
\end{equation}
Since the temperature of the domain decreases over time, it is necessary to consider various scenarios, taking into account all the processes possible under the given conditions:\\
$\mathbf{T\in[2.83\times 10^8,10^7]}\text{K}$\\
The following processes are possible during this period:\\
$\cdot$pair product during the interaction of annihilation and thermal photons and the further annihilation of an electron with a positron; \\
$\cdot$multiple Compton scattering of an annihilation photon on a positron;\\
The decay of a negatively charged muon and the further annihilation of an electron with a positron.\\  
In this case, the formation of pairs can be considered as the leading process affecting the mean free path of the annihilation photon. Total interaction cross section for an annihilation photon at a given temperature:
\begin{equation}
    \sigma \approx\sigma_{pp}\approx2,1\times10^{-28} \text{cm}^2.
\end{equation}
$\mathbf{T\leq10^7}\text{K}$.
Starting from the moment when the temperature of the domain becomes equal to $T=10^7$K ($t=10^6$c), the formation of positron-electron pairs becomes unlikely even taking into account the high-energy tail in the Planck distribution. In this case, two processes will take place inside the domain:\\
$\cdot$multiple Compton scattering;\\
$\cdot$decay of a negatively charged muon and the further annihilation of an electron with a positron.\\  
At the same time, Compton scattering on a positron should be considered the leading process affecting the mean free path of an annihilation photon. The cross section for the annihilation photon in this case is:
\begin{equation}
\sigma=\sigma_k\approx8.4\times10^{-22}\text{cm}^2.
\end{equation}
\subsection{Penetration depth of photons
} \label{4}
The depth of photon penetration into the domain is determined by their mean free path, which can be calculated:
\begin{equation}
    \lambda={\frac{1}{n\sigma}},
\end{equation}
where\\
n-thermal photon/positron density,\\
$\sigma$-cross section.\\
In case of pair production:
\begin{equation}
    \lambda_{pp}=\frac{1}{T^3\sigma_{pp}}
\end{equation}
\begin{equation}
    2\times10^2\text{cm}\leq\lambda_{pp}\leq5\times10^6\text{cm}.
\end{equation}
In case of Compton scattering:
\begin{equation}
    n_p=\frac{\rho Z}{m_p},
\end{equation}
where Z=1-the average number of electrons per nucleon for a domain consisting mainly of antihydrogen and antihelium.\\
Then, penetration depth for the multiple Compton scattering:
\begin{equation}
    \lambda_k =\frac{1}{T^3\eta Z\sigma_{eff}} 
\end{equation}
For $T=10^7$K, $\sigma_{eff}=8.4\times10^{-22}\text{ cm}^2$:\\
$\eta=3\times10^{-12}:\lambda_k=4\times10^{11}$cm;\\
$\eta=1\times10^{-6}:\lambda_k=3\times10^{5}$cm.\\
The obtained photon penetration depth can be compared with the domain size. Assuming spherical symmetry, we write the following inequality:
\begin{equation}
    \lambda<R.
\end{equation}
Since the formation of positron-electron pairs is possible only at temperatures of $T\geq10^7$K, we determine the domain size in the time period $1,25\times 10^3\leq t\leq 10^6$c:\\
$R_{min}\approx10^{13}$cm;\\
$R_{max}\approx10^{17}$cm.\\
Comparing with the possible range for the path length of $10\leq\lambda_{pp}\leq 10^5$cm, we can conclude that at any given time in the considered range, the penetration depth of annihilation photons is significantly less than the domain size. Thus, due to the high concentration of thermal photons in the radiation era, the interaction of photons will occur close to the domain boundary. At the same time, there will always be an area within the domain where the formation of pairs will not occur.\\
Since Compton scattering is possible at any temperature, the inequality relating the path length to the domain size will look like this:
\begin{equation}
    \frac{1}{T^3\eta Z\sigma_{eff}}<R
\end{equation}
Considering the conditions under which Compton scattering is the dominant process and he expression found for the domain size:
$$\begin{cases}
    t>10^6\text{c} \\
t<\frac{Z10^{20}\sigma_{eff} \eta^{\frac{2}{3}}M^{\frac{1}{3}}}{m_p^{\frac{1}{3}}}\approx\eta^\frac{2}{3}M^{\frac{1}{3}}\times7\times10^6 \text{c}
\end{cases}$$
\section{Conclusion} \label{5}
In the course of the work, the main processes occurring at the boundary of the antimatter domain and inside it were considered.\\
The formation of electron-positron pairs during the interaction of a thermal photon with an annihilation photon should be considered a key process affecting the depth of photon penetration because of the concentration of thermal photons inside the domain during the radiation era. The formation of such pairs leads to the annihilation of positrons with electrons inside the domain. In addition, the decay of a negatively charged pion as an electron source should also be taken into account.\\ 
At $T\leq10^7$K, the leading process occurring inside the domain is multiple Compton scattering of annihilation photons by positrons, as a result of which the photon energy will decrease.\\
For each of the processes described above, the interaction cross-section and the photon penetration depth were estimated. The values obtained were compared with the domain size at the corresponding time. According to the calculated values, it can be concluded that at $T\geq10^7$K, the depth of photon penetration into the domain is significantly less than its size. In the case of Compton scattering, the path length of the annihilation photons will not exceed the size of the domain only in a limited time range, depending on the mass of the domain and its density.Such an assessment makes it possible to determine an nonhomogeneous region of the antimatter domain, within which various processes will occur that affect its chemical structure.\\
In the future, a more detailed description of the processes occurring inside the domain is planned, as they will make changes to the chemical structure of the domain. Based on the results obtained, the most accurate assessment of a homogeneous region is possible, which is not subject to changes as a result of processes occurring at the boundary of the domain with the horizon. The main task of further work is to study the evolution of the domain over time.
\section*{Acknowledgements}
The work of M. K. was performed in Southern Federal University with financial support of grant of Russian Science Foundation № 25-07-IF.

\end{document}